\documentclass[prd,superscriptaddress,floatfix,nofootinbib,twocolumn,aps,nobibnotes]{revtex4-1}
\usepackage{graphicx}
\usepackage{dcolumn}
\usepackage{bm}
\usepackage{xcolor}
\usepackage{amsmath,amssymb,txfonts}
\usepackage{ulem}
\usepackage{siunitx}
\usepackage{soul}
\usepackage[english]{babel}
\usepackage[colorlinks,linkcolor=red,citecolor=blue,urlcolor=red]{hyperref}

\begin{document}

\title{Searching for Chameleon Dark Energy with Mechanical Systems}
\author{J. Betz}
\affiliation{Department of Physics and Astronomy, University of Delaware, Newark, DE 19716, USA}

\author{J. Manley}
\affiliation{Department of Electrical and Computer Engineering, University of Delaware, Newark, DE 19716, USA}

\author{E. M. Wright}
\affiliation{Wyant College of Optical Sciences, University of Arizona, Tucson, AZ 85721, USA}

\author{D. Grin}
\affiliation{Department of Physics and Astronomy, Haverford College, Haverford, PA 19041, USA}

\author{S. Singh}
\email{swatis@udel.edu}
\affiliation{Department of Electrical and Computer Engineering, University of Delaware, Newark, DE 19716, USA}
\affiliation{Department of Physics and Astronomy, University of Delaware, Newark, DE 19716, USA}

\date{\today}

\begin{abstract}

A light scalar field framework of dark energy, sometimes referred to as quintessence, introduces a fifth force between normal matter objects. Screening mechanisms, such as the chameleon model, allow the scalar field to be almost massless on cosmological scales while simultaneously evading laboratory constraints. We explore the ability of mechanical systems available in the near term to directly detect the fifth force associated with chameleon dark energy. We provide analytical expressions for the weakest accessible chameleon model parameters in terms of experimentally tunable variables and apply our analysis to two mechanical systems: levitated microspheres and torsion balances, showing that the current generation of these experiments have the sensitivity to rule out a significant portion of the proposed chameleon parameter space. We also indicate regions of theoretically well-motivated chameleon parameter space to guide future experimental work.


\end{abstract}

\maketitle

{\it Introduction. ---} Multiple cosmological measurements \cite{Riess_1998, Perlmutter_1999,Aghanim_2020, Alam_2020} indicate the presence of a novel negative-pressure fluid with a constant energy density that dominates the energy budget of the Universe during the present epoch \cite{Steinhardt2006}. There is no consensus on the theoretical framework for the composition, properties or production mechanism of this fluid, known as dark energy (DE), which could be responsible for the observed accelerated expansion of the Universe \cite{Caldwell:2009ix,Brax_2017}. Theoretical approaches for building a DE framework typically involve the introduction of light scalar fields or modifications to General Relativity \cite{Steinhardt2006,Caldwell:2009ix,Joyce:2016vqv,Brax_2017}. In order to explain the observed cosmic acceleration, both scenarios must contend with new degrees of freedom mediating long-range forces between Standard Model (SM) particles. A variety of experiments have placed tight constraints on the long-range fifth force between SM particles due to such scalar fields. These constraints, however, can be evaded by a class of theories known as `screened-scalar' models \cite{Khoury:2003rn,Khoury_2004,Brax_2004,Khoury:2010xi,Hinterbichler:2011ca,Brax:2013doa,Burrage:2014uwa,Brax_2017,Burrage_2018,Burrage:2020bxp} such as the chameleon model.

In the chameleon model, the effective mass of the chameleon field is dynamically modified by terms that depend on the local matter density \cite{Khoury_2004,Khoury_Symmetron,Brax_2017}.
Because of this mechanism, the scalar field can remain light on cosmic scales, allowing it to behave as vacuum energy, yet heavy in laboratory environments, where screening suppresses the fifth force, allowing it to evade detection. The strong dependence on the local matter density causes the chameleon field and corresponding force between two objects to be extremely sensitive to the local geometry and surrounding environment. 

Supplementing observational constraints on screened-scalar models \cite{Brax:2005ew,Bean:2007nx,Bean:2007ny,Brax:2011pk,Davis:2011pj,Jain_2013,Brax:2017wcj,Vikram_2018,OHare:2018ayv,Tamosiunas:2021kth,Lagos:2020mzy}, the best laboratory constraints on the chameleon model come from the E{\"o}t-Wash torsion pendulum \cite{EotWash_2012,Kapner_2007} and atom interferometry experiments \cite{Jaffe_2017}. The E{\"o}t-Wash torsion pendulum is able to position large source and test masses with micron separations, providing excellent constraints on scalar field mediated forces. However, large masses make it difficult to probe stronger chameleon coupling strengths because of screening. Atom interferometry experiments can access a complementary chameleon parameter space using smaller test masses. However, they are limited to stronger coupling strengths because the small masses result in a correspondingly small chameleon force. These limitations leave a large area of the chameleon parameter space unprobed. Precision measurement experiments with intermediate size masses are uniquely suited to fill this gap in the chameleon constraint space.

\begin{figure}
    \centering
    \includegraphics[]{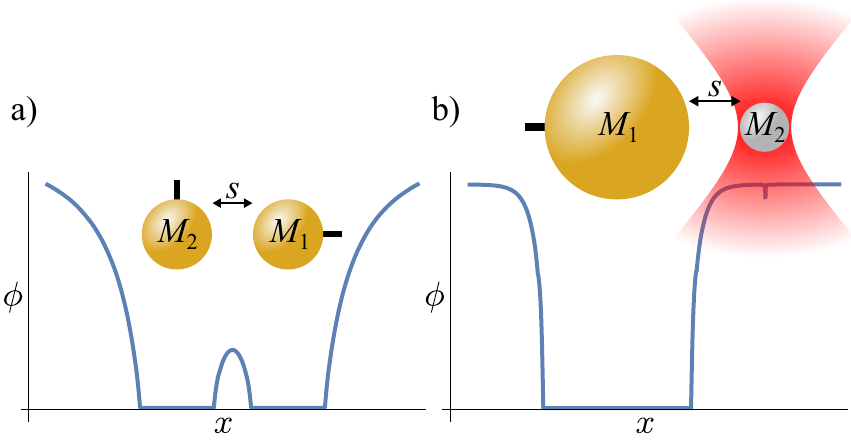}
    \caption{Schematic of mechanical systems considered here to detect the chameleon mediated force between matter. In both cases, $M_1$ is the gold source mass, while $M_2$ is the mechanically compliant test mass. We consider (a) torsion balance and (b) optically levitated microspheres as sensors of chameleon DE. The blue curves show the two-body chameleon field profile. Notice, the levitated microsphere is enlarged to show detail.}
    \label{fig:diagrams}
\end{figure}

In this letter, we present a widely applicable theoretical treatment of mechanical systems as sensors of the fifth force associated with chameleon DE. Considering spherically symmetric masses, we provide analytical expressions for the weakest accessible chameleon field phenomenology parameters (such as the self-interaction strength and coupling to normal matter) in terms of experimental parameters such as size, distance between spheres, and the minimum detectable force. Our expressions provide a straightforward pathway to estimate the performance of existing devices and scaling arguments useful for design considerations of future devices, without the need for numerically intensive solutions.

We then discuss two classes of spherical mechanical systems: levitated microspheres and Cavendish-style torsion balances. Both systems have demonstrated exceptional sensitivity to weak forces \cite{Montoya_2021,Ranjit_2016, westphal_2021} and there has been some work considering microspheres as potential probes for scalar and screened-scalar fields \cite{Rider_2016, geraci2010short, Blakemore_2021, Qvarfort_2021}. Using a novel analytical treatment allowing us to consider systems with larger masses, we show that for both mechanical systems, the current generation of experiments have the sensitivity to put new constraints in a region of interest to cosmology, possibly ruling out chameleons as DE. We also present the region of phenomenologically motivated chameleon DE parameter space that future experiments could be optimized to target.

{\it Chameleon Model. ---} The chameleon force between two objects of masses $M_1$ and $M_2$ can be approximated by \cite{Hui_2009,Burrage_2015}
\begin{equation} \label{eq:force}
F_\text{cham}(x) = 2 \ \alpha \frac{G M_1 M_2}{x^2}\lambda_1 \ \lambda_2 \left(\frac{M_\text{p}}{M}\right)^2,
\end{equation}
where $M$ is the chameleon-matter coupling, $M_\text{p}$ is the reduced Planck mass, $x$ is the center of mass separation distance, $\lambda_{1,2}$ are the chameleon screening factors associated with each object, and $\alpha$ is a dimensionless factor (see below).
The screening factors are given by \cite{Burrage_2015}
\begin{equation}
\label{eq:screening}
\lambda_i = 
\left\{\begin{array}{cc}
1, &  \rho_i R_i^2 < 3 M \phi_\text{bg}, \\
\approx \dfrac{3 M \phi_\text{bg}}{\rho_i R_i^2}, & \rho_i R_i^2 > 3 M \phi_\text{bg},
\end{array}\right.
\end{equation}
where $\phi_\text{bg} = \left(n M \Lambda^{4+n}/\rho_\text{bg}\right)^{1/(n+1)}$ is the background value of the chameleon field, $R_i$ and $\rho_i$ are the radius and density of the object. The background field value, $\phi_\text{bg}$, depends on the background density, $\rho_\text{bg}$, the chameleon self-interaction coupling, $\Lambda$, and the power-law index, $n$. The coupling parameters $M$, $\Lambda$, and $n$, which come from the chameleon equation of motion,
\begin{equation}\label{eq:eom}
\nabla^2\phi = -\frac{\Lambda^{4+n}}{\phi^{n+1}}+\frac{\rho}{M},
\end{equation}
are the three independent parameters of the chameleon model.
As Eq. (\ref{eq:screening}) demonstrates, with all other parameters held fixed, a larger or more dense object will be screened more relative to a smaller, less dense one. Additionally, increasing the density surrounding an object will cause more screening. It is this screening mechanism which prevents observations of the chameleon force between macroscopic objects.

The chameleon force between two objects derived in Refs. \cite{Hui_2009,Burrage_2015,Qvarfort_2021} assumes a mass hierarchy between the source and test mass such that the chameleon field sourced by the test mass can be treated as a perturbation to the source mass field. In this regime the two-body chameleon field can be approximated by $\phi(\mathbf{x}) \approx \phi_1(\mathbf{x} - \mathbf{x}_1) + \phi_2(\mathbf{x} - \mathbf{x}_2) - \phi_{bg}$ where $\phi_{1,2}$ are the one-body field solutions for the source and test mass. We found that for both masses of a similar scale, this approximation breaks down as the inherent nonlinearity in the chameleon equation of motion prevents the two-body field from being approximated as a superposition of the one-body solutions. However, this similar scale regime is of interest because, as shown below, experiments operating in this regime can be used to set new bounds on the chameleon parameter space. 

Starting with a multiplicative ansatz for the two-body field solution, $\phi(\mathbf{x}) \approx \phi_1(\mathbf{x} - \mathbf{x}_1) \ \phi_2(\mathbf{x} - \mathbf{x}_2) /\phi_\text{bg}$, we have derived an expression for the chameleon force between two spherical objects without the limitation of a mass hierarchy between the spheres (see supplemental material \cite{SI}). Taking the same approximations used in Ref. \cite{Burrage_2015}, we found a force expression that matched the previous result ($\alpha = 1$) \cite{Hui_2009,Burrage_2015,Qvarfort_2021} except in the regime where both spheres are strongly perturbing. In this case, our force is smaller by a factor of $\alpha = 1/6$. When the one-body field solutions overlap significantly, the additive ansatz can lead to negative (and thus unphysical) solutions even in the weakly perturbing regime. The multiplicative ansatz, by construction, is never negative. For the experimental systems considered here, we found the multiplicative ansatz a good approximation to the numerical two-body field solution in the strongly perturbing regime. In the weakly perturbing regime, the multiplicative ansatz is not as accurate a field solution as the additive ansatz solution (see supplemental material \cite{SI}), but still provides an accurate force calculation, as it agrees with the previous results. A detailed analysis of the applicability of the multiplicative ansatz and corresponding force derivation will be elaborated in a future paper.

\begin{figure}
    \centering
    \includegraphics[]{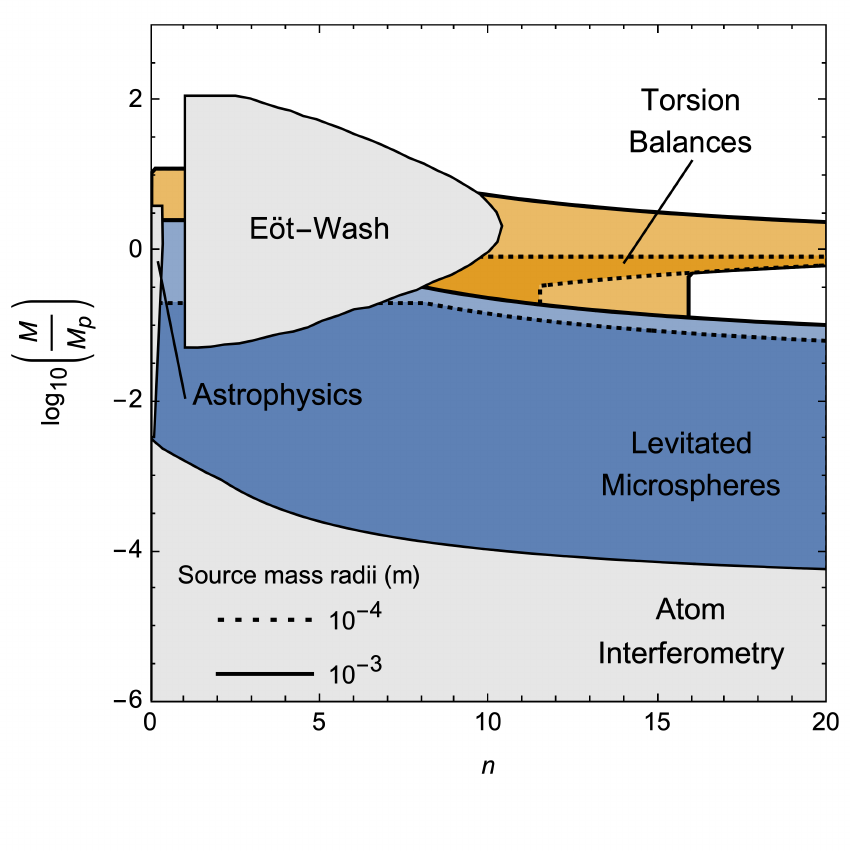}
    \caption{Estimated constraints from levitated microspheres (blue) and torsion balances (orange) as chameleon dark energy detectors, where the chameleon self-coupling is fixed at $\Lambda = \Lambda_\text{DE}$. In this plot, we consider a microsphere system with $R_1=5\,\mu$m, $s=0.5$ mm, and $F_\text{min} = 8.5 \times 10^{-22}$ N, and a torsion balance system with $R_1=0.1$ mm, $s=0.6$ mm, and $F_\text{min} = 2 \times 10^{-18}$ N. For both systems, the solid (dashed) black line corresponds to source a mass with $R_1=1$ mm ($0.1$ mm).
    The multiplicative ansatz and force expression are used to generate the torsion balance curves in the region where the additive ansatz breaks down. 
    Existing constraints from Ref. \cite{Burrage_2018}, which include results from Refs. \cite{EotWash_2012,Jaffe_2017,Jain_2013,Vikram_2018}, are shown in gray. See the supplemental material \cite{SI} for a qualitative description of chameleon constraint plots.}
    \label{fig:n>1constraints}
\end{figure}

First, we consider chameleon DE models by fixing $\Lambda = \Lambda_\text{DE} \sim 2.4 \times 10^{-3} \text{ eV}$ \cite{Aghanim_2020}. The free parameters are the chameleon-matter coupling, $M$, and the power-law index, $n$. 


For laboratory scale experiments with a minimum force sensitivity, $F_\text{min}$, setting $F_\text{cham} = F_\text{min}$ yields an analytic expression for the maximum $M$ value that can be probed by the experiment,
\begin{equation}\label{eq:Mmax}
\frac{M_\text{max}}{M_\text{p}} = \left(\frac{32 \pi^2}{9} \frac{G \rho_1 \rho_2}{F_\text{min}} \frac{R_1^3 R_2^3}{(s+R_1+R_2)^2}\right)^{1/2}.
\end{equation}
All parameters on the right side of the equation are in SI units including the surface separation distance, $s$.
The minimum detectable force,  $F_\text{min}$, may depend on the radius of the test mass. Where $M = M_\text{max}$, both the source and test mass are unscreened ($\lambda_{1,2} = 1$) and the chameleon force is independent of $n$ and $\Lambda$. Additionally, the maximum $n$ value can be found numerically by solving the following equation,
\begin{multline}\label{eq:nmax}
F_\text{min} = 4 \pi \xi^2 \frac{R_1 R_2}{\left(s+R_1+R_2\right)^2} \left(\frac{1}{\hbar c}\right)^{\dfrac{n_\text{max}+6}{n_\text{max}+2}} \\
\times \left[n_\text{max}(n_\text{max}+1)(\Lambda_\text{DE})^{4+n_\text{max}}L^2\right]^{\dfrac{2}{n_\text{max}+2}}.
\end{multline}
Here, all parameters are in SI units. The dimensionless factor, $\xi$, is a constant which characterizes the geometry of the vacuum chamber. For a spherical vacuum chamber of radius, $L$, $\xi = 0.55 - 0.68$, which was found numerically in Ref. \cite{Hamilton_2015} in addition to other vacuum chamber geometries.  

More generally than dark energy (ie even if $\Lambda\neq\Lambda_\text{DE}$), chameleon screening could hide scalar fields appearing in string-inspired scenarios beyond the Standard Model \cite{Damour:1994zq,Brax:2010gi}, motivating laboratory searches. Focusing on $n\geq1$, analytic expressions can be found for the maximum $M$ and minimum $\Lambda$ values that can be probed by a particular experiment. $M_\text{max}$ is given by Eq. (\ref{eq:Mmax}) and $\Lambda_\text{min}$ by,
\begin{equation}\label{eq:Lmin}
\Lambda_\text{min} = \left(\frac{F_\text{min}}{4\pi\xi^2} \frac{(s+R_1+R_2)^2}{ R_1 R_2}\right)^{3/10} \left(\frac{1}{L}\right)^{2/5} \frac{(\hbar c)^{7/10}}{1.6\times10^{-19}}.
\end{equation}
Here, $\Lambda$ is in eV, and all parameters on the right side of the equation are in SI units. 

\begin{figure}
    \centering
    \includegraphics[]{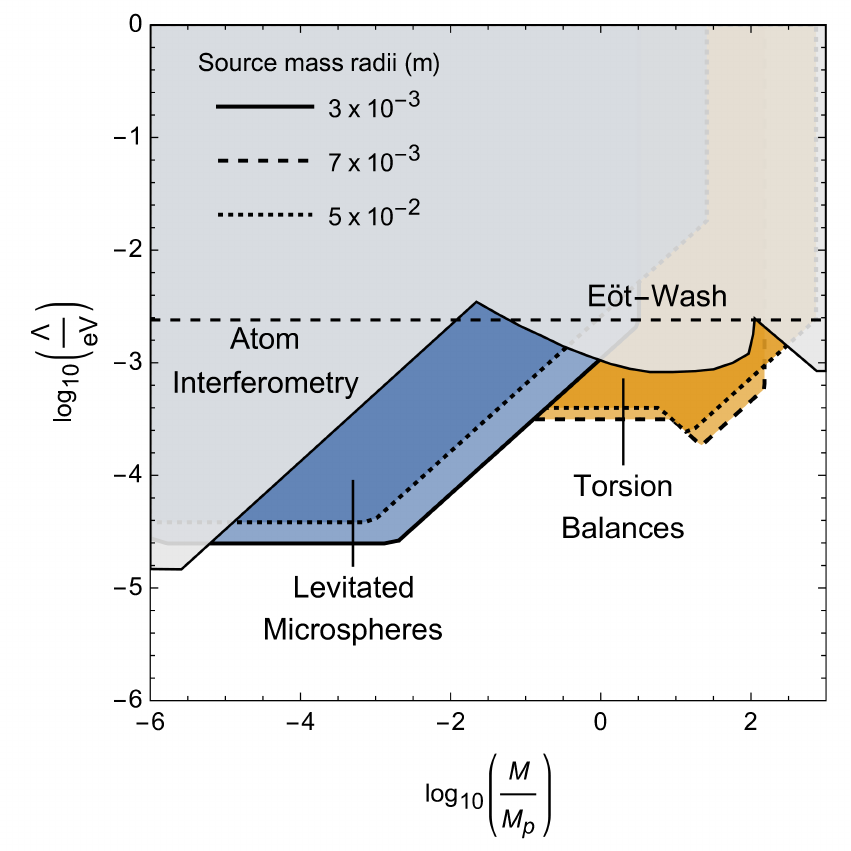}
    \caption{Estimated constraints from levitated microspheres (blue) and torsion balances (orange) as chameleon dark energy detectors, where the power-law index is fixed at $n = 1$. In this plot, we consider a microsphere system with $R_2=5\,\mu$m, $s=3$ mm, and $F_\text{min} = 8.5 \times 10^{-22}$ N, and a torsion balance system with $R_2=1$ mm, $s=6$ mm, and $F_\text{min} = 5 \times 10^{-17}$ N. For torsion balances, the dashed (dotted) black line corresponds to a source mass with $R_1=7$ mm ($5$ cm).
    The multiplicative ansatz and force expression are used to generate the torsion balance curves in the region where the additive ansatz breaks down. 
    For microspheres, the solid (dashed) black line corresponds to a source mass with $R_1=3$ mm ($7$ mm). 
    The horizontal black dashed line indicates the dark energy scale $\Lambda = \Lambda_\text{DE}$. Existing constraints from Ref. \cite{Burrage_2018}, which include results from Refs. \cite{EotWash_2012,Jaffe_2017}, are shown in gray. }
    \label{fig:n1constraints}
\end{figure}

{\it Experiment. ---} Figure \ref{fig:diagrams} illustrates a simplified schematic for a general fifth force experiment. The force between two masses is inferred by measuring the position of a test mass $M_2$ (modeled as a harmonic oscillator) with radius $R_2$, in response to the force exerted on it by a source mass $M_1$ with radius $R_1$. The position of the source mass is made to oscillate, resulting in an oscillating force signal $F(t)$ on the test mass. One way to accomplish this would be to have the source mass rotate in a circle around the test mass. The oscillating force signal is necessary to avoid low frequency noise, and the driving frequency may be chosen near the oscillator's resonance frequency. With resonant amplification and sufficiently low-noise displacement readout, a force measurement will be limited by thermomechanical noise in the oscillator. 

A harmonic oscillator with dissipation $\Gamma_0$ and effective mass $m_0$, operating at a finite temperature $T$, will have a thermal force noise spectrum $S_{FF}^\text{th}(f)=4 k_\text{B} T m_0 \Gamma_0$. We have approximated the force signal as monochromatic, and the signal to noise ratio for a coherent signal over stochastic noise can be improved by averaging down the variance in the noise floor over a longer measurement time $\tau$, so that the thermally-limited minimum detectable force is
\begin{equation}\label{eq:fmin}
    F_\text{min}\approx \sqrt{2\, S_{FF}^\text{th}/\tau}=\sqrt{8  k_\text{B} T m_0 \Gamma_0/\tau}.
\end{equation}
While this expression holds irrespective of the source or test mass geometries, we consider the specific case of spherical masses, for which an analytical expression for the force due to a chameleon field is provided by Eq. (\ref{eq:force}). 

Multiple existing force sensors with spherical test masses are capable of performing thermally-limited force measuremnts, achieving low enough sensitivities to probe chameleon dark energy models. In Figs. \ref{fig:n>1constraints} and \ref{fig:n1constraints}, we plot the estimated constraints that can be set on chameleon models by two classes of sensors: optically levitated microspheres and torsion balances. 

The blue shaded regions in Figs. \ref{fig:n>1constraints} and \ref{fig:n1constraints} demonstrate the estimated reach of levitated microspheres, where the test mass is a silica sphere with radius $R_2=5 \, \mu\text{m}$ that is confined within a harmonic potential via optical trapping. Such sensors operating in high vacuum have been shown to achieve attonewton $/\sqrt{\rm Hz}$ force sensitivities~\cite{monteiro2020force}. For simplicity, we assume gas damping (due to collisions with the surrounding gas molecules) as the dominant dissipation mechanism giving rise to thermomechanical noise. In a vacuum chamber at pressure $10^{-6}$ mbar and temperature $T=300$ K, we estimate $\Gamma_0 \approx 10^{-4}\, {\rm s}^{-1}$ from an expression in Ref. \cite{geraci2010short} for gas damping. This yields a thermally-limited force sensitivity of $ 10^{-18} \, \rm N/\sqrt{\rm Hz}$, and a minimum detectable force of $ 8.5 \times 10^{-22}$ N for a $\tau=2$ month measurement. 

To supplement the potential constraints from levitated microspheres, we also consider torsion balances, whose larger test masses enable them to probe weaker chameleon-matter couplings (larger $M$). A simple Cavendish-style torsion balance \cite{westphal_2021} consists of equally-sized spheres, connected by a rod of negligible mass, which is suspended at its center by a torsion fiber and placed in a vacuum chamber. One of the spheres serves as the test mass $M_2$, while the balance as a whole forms a harmonic oscillator with effective mass $m_0\approx 2 M_2$ and quality factor $Q_0=2\pi f_0 / \Gamma_0$. Westphal et al. \cite{westphal_2021} have demonstrated that such torsion balances have the ability to make measurements near the thermal limit, achieving piconewton $/\sqrt{\rm Hz}$ force sensitivity. In Figs. \ref{fig:n>1constraints} and \ref{fig:n1constraints}, the orange regions correspond to the estimated constraints that can be set on chameleon models by torsion balances operating at pressure $10^{-6}$ mbar and temperature $T=300$ K, assuming a quality factor of $Q_0=10$ and a torsional resonance frequency of $f_0=5$ mHz. In Fig. \ref{fig:n>1constraints} (\ref{fig:n1constraints}) we consider a gold test mass $R_2=0.1$ mm ($R_2=1$ mm), achieving a thermally-limited force sensitivity of $ 3 \times 10^{-15}\, \rm N/\sqrt{\rm Hz}$ ($ 9 \times 10^{-14}\, \rm N/\sqrt{\rm Hz}$) and a minimum detectable force of $ 2\times 10^{-18}$ N ($5\times 10^{-17}$ N) for a $\tau=2$ month measurement. 

For both systems, we assume gold source masses, with various radii labeled in Figs. \ref{fig:n>1constraints} and \ref{fig:n1constraints}. The surface separation distances $s$ between the test and source masses (see figure captions for values) are chosen such that Casimir forces are negligible relative to $F_\text{min}$. 

A common component in torsion balance experiments is an electromagnetic shield placed between the source and test masses \cite{Kapner_2007,westphal_2021}. These shields can also be used to control Casimir forces which are relevant for the geometries proposed here \cite{Bimonte_2018,Qvarfort_2021}. However, this shield will introduce further screening of the chameleon force which can be estimated analytically \cite{EotWash_2012} or calculated numerically. 

These experimental limitations can be overcome through better force sensitivity which will also allow probing of smaller $\Lambda$, and larger $M$. For fixed $s$ and $R_2$, $M_\text{max}$ is a monotonically increasing function of $R_1$, indicating that larger source masses are needed to probe weaker chameleon-matter coupling strengths. However, when fixing the same parameters, $\Lambda_\text{min}$ can be minimized by a particular choice for $R_1$. This behavior is illustrated in Fig. \ref{fig:n1constraints}; $\Lambda_\text{min}$ has been optimized for the light blue and light orange regions by varying $R_1$ with fixed $s$ and $R_2$. Increasing $R_1$ from the optimal value probes larger $M$ at the expense of larger $\Lambda$.

\begin{figure}
    \centering
    \includegraphics{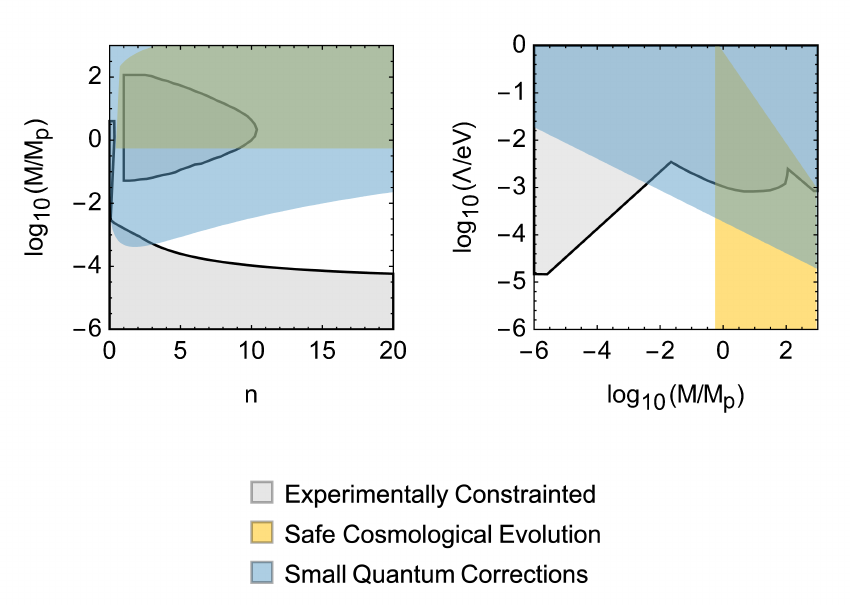}
    \caption{The parameter space for both $n = 1$ and $n > 1$ chameleon models considered here is limited by cosmological \cite{Erickcek_2014,Brax_2004} as well as quantum field theory analyses \cite{Upadhye_2012}. The green regions indicate where the chameleon model has safe cosmological evolution and small quantum corrections and should be the aim of future experimental searches. The gray regions indicate existing constraints (shown in Figs. \ref{fig:n>1constraints} and \ref{fig:n1constraints}). See Ref. \cite{SI} for region definitions. }
    \label{fig:safespace}
\end{figure}

It is important to note that the geometries proposed here do not set new constraints for unscreened scalar fields (Yukawa 5th forces). Unscreened scalar fields can evade experimental detection only through their large masses and subsequent exponential suppression which results in an extremely short range force. Thus, in order to detect these scalars, large masses at very small separation distances are required. Screened scalars, on the other hand, are generally very light in low density regions and can therefore have a significantly longer range force which scales as $1/r^2$. Rather than small separation distances, the key in designing experiments to search for screened scalars is to prevent screening of the source and test masses. This can be controlled by various parameters such as density, geometry, and size of the source and test masses, as well as the vacuum chamber size and pressure.

{\it Cosmological and Naturalness Constraints. ---} Beyond the addition of a new force, introducing a new particle to the Standard Model can impact cosmological models and physics in the early universe. For example, in Ref. \cite{Erickcek_2014} it was shown that for sufficiently strong matter-coupling (small $M$), the chameleon is kicked down its potential as other particle species become non-relativistic. As the field rebounds off the steep side of its potential, extremely high energy modes are generated - invalidating the treatment of the chameleon as a low-energy effective field theory and potentially disrupting Big Bang nucleosynthesis. For chameleons with matter couplings weaker than, $\log_{10}(M/M_\text{p}) \gtrsim -0.26 $, this breakdown can be avoided only for certain initial conditions \cite{Erickcek_2014}. On the other hand, requiring the chameleon to behave as vacuum energy in the current epoch places a bound on the weakest allowable coupling. In Ref. \cite{Brax_2004} it was shown that the chameleon will follow an attractor solution provided the condition $m^2/H^2 \gg 1$, where $m$ is the chameleon mass and $H$ is the Hubble parameter. For fixed $n$ and $\Lambda$, this places a bound on the maximum value of $M$. Thus, for the chameleon model as written, only a particular region of parameter space is cosmologically well motivated with small quantum corrections. The range of parameter space satisfying both these constraints is indicated by the green region in Fig. \ref{fig:safespace}. However, modified chameleon models such as the Dirac-Born-Infeld (DBI) chameleon \cite{Burrage:2014uwa,Padilla_2016} may be feasible outside of this restricted parameter space.

While we omitted projected constraints from Figs. \ref{fig:n>1constraints} and \ref{fig:n1constraints} for clarity, there is some overlap with the regimes of interest shown in Fig. \ref{fig:safespace}. Both systems, with current sensitivities, can be optimized to probe more of the green region. Additionally, having a larger test mass would enable us to probe deeper into the green region in Fig. \ref{fig:safespace}. This may be accomplished via magnetic levitation of the test mass \cite{Timberlake_2019, Xiong:2021gby, Lewandowski_2021, Brown_2021}, albeit with design considerations to avoid technical noise and screening due to the nearby matter required to create such traps.

{\it Conclusion. ---} Current generation mechanical systems have the sensitivity to rule out significant portions of chameleon parameter space and cast doubt on the feasibility of chameleon dark energy. 
For inverse power-law models, only weakly coupled (gravitational strength) chameleons have viable early cosmological evolution \cite{Erickcek_2014}. On the other hand, it has been shown that the $n = -4$ model is cosmologically safe \cite{Miller_2016}. Future work will extend predicted constraints for these mechanical systems to negative $n$ chameleon models and other screened scalar fields. We will also explore using mechanical systems with reduced geometries (such as disks, membranes or strings), as DE detectors as the chameleon force may be enhanced between non-spherical objects \cite{Burrage_2017}.

Beyond DE and modified gravity theories, screening mechanisms can also be utilized to hide scalar fields coming from string theory. Mechanical systems are particularly well suited to search for such screened-scalar fields. Size, geometry and material flexibility coupled with excellent force sensitivity makes them an ideal experimental platform for optimized searches for a variety of screening mechanisms.

We would like to thank Brian LaRocca and Mark Mirotznik for assistance with numerical simulations, and Natalie Schmidt for assistance with figures. We also acknowledge helpful discussions with Andy Geraci, Jiyan Sang and Ryan Petery. We thank P. Meystre for feedback on the manuscript. This work is supported by the National Science Foundation Grant No. PHY-2112846, CAREER Grant No. PHY-2047707 and the Provost’s Office at Haverford College.

\bibliographystyle{apsrev4-1}
\bibliography{main}

\end{document}


\title{Supplemental Material for \\ ``Searching for Chameleon Dark Energy with Mechanical Systems"}

\author{J. Betz}
\affiliation{Department of Physics and Astronomy, University of Delaware, Newark, DE 19716, USA}

\author{J. Manley}
\affiliation{Department of Electrical and Computer Engineering, University of Delaware, Newark, DE 19716, USA}

\author{E. M. Wright}
\affiliation{Wyant College of Optical Sciences, University of Arizona, Tucson, AZ 85721, USA}

\author{D. Grin}
\affiliation{Department of Physics and Astronomy, Haverford College, Haverford, PA 19041, USA}

\author{S. Singh}
\email{swatis@udel.edu}
\affiliation{Department of Electrical and Computer Engineering, University of Delaware, Newark, DE 19716, USA}
\affiliation{Department of Physics and Astronomy, University of Delaware, Newark, DE 19716, USA}

\email{swatis@udel.edu}
\date{\today}

\maketitle

\numberwithin{equation}{section}

\tableofcontents

\section{Interpreting Chameleon Constraint Plots}\label{sec:constraints}

The chameleon force between two spheres is given in the main text by
\begin{equation}\label{eq:force}
F_\text{cham}(x) \approx 2 \ \alpha \frac{G M_1 M_2}{x^2}\lambda_1 \lambda_2 \left(\frac{M_\text{p}}{M}\right)^2,
\end{equation}
where $x$ is the center of mass separation distance. 
%
\begin{figure}
    \centering
    \includegraphics[]{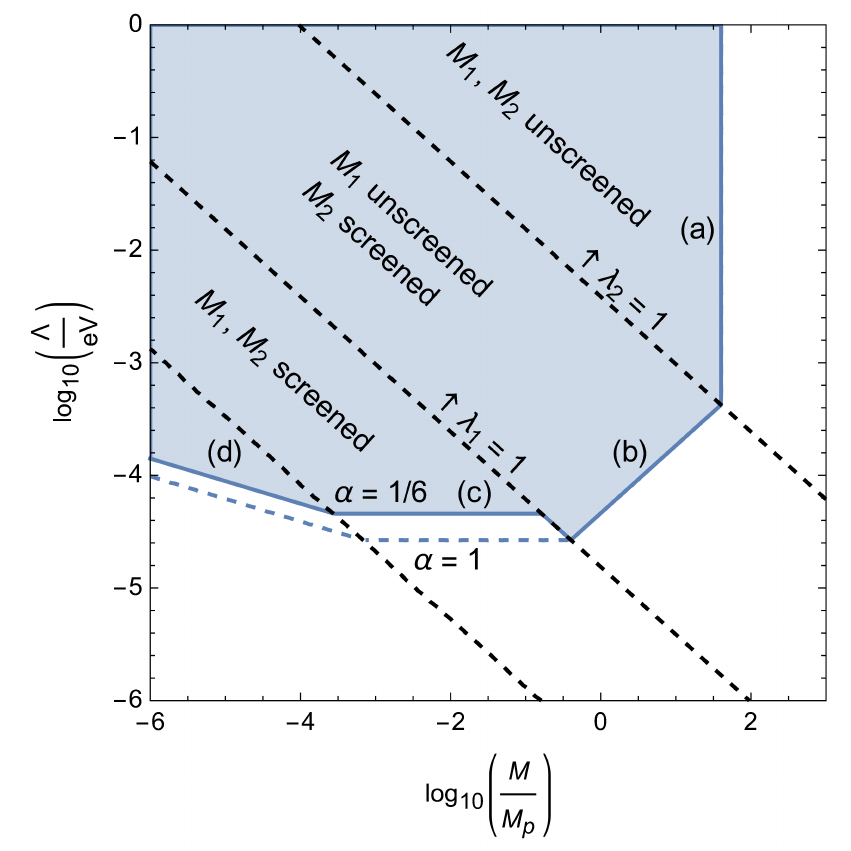}
    \caption{Fixed $n$ chameleon constraint plot illustrating the behavior of the chameleon force as the source and test mass become screened. The shaded region indicates $F_\text{cham} \geq F_\text{min}$.}
    \label{fig:n1plot}
\end{figure}
\begin{figure}
    \centering
    \includegraphics[]{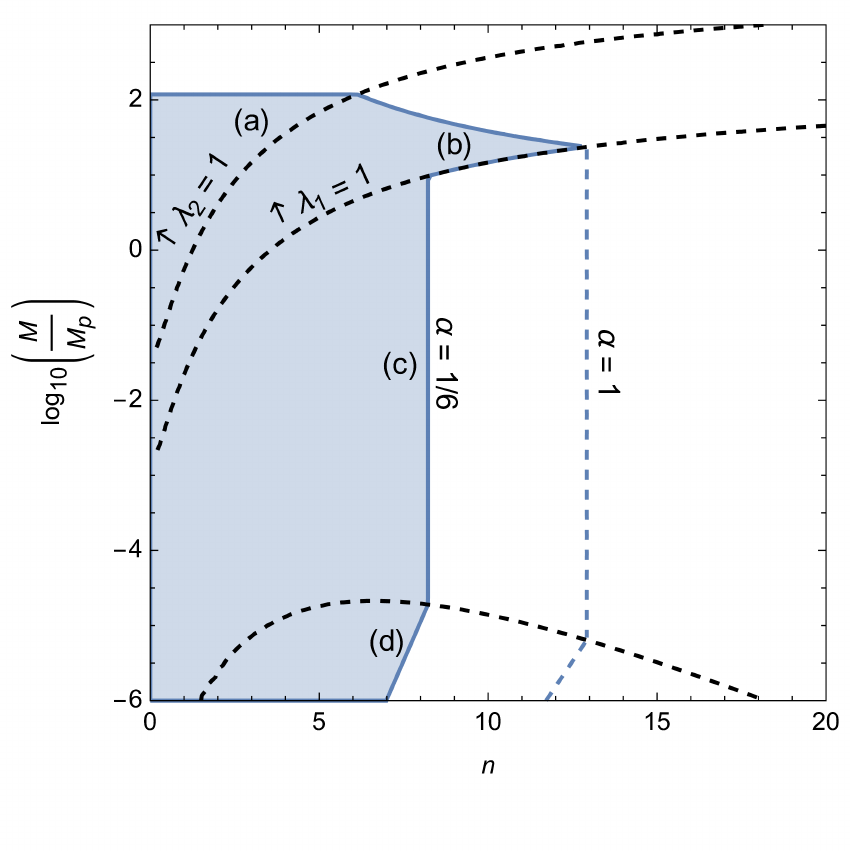}
    \caption{Fixed $\Lambda$ chameleon constraint plot illustrating the behavior of the chameleon force as the source and test mass become screened. The shaded region indicates $F_\text{cham} \geq F_\text{min}$.}
    \label{fig:n>1plot}
\end{figure}
The chameleon force has three to four regions of distinct behavior as $M$, $\Lambda$, and $n$ are varied as illustrated in Fig.~\ref{fig:n1plot} and \ref{fig:n>1plot}. This behavior is caused by the screening factors $\lambda_{1,2}$, where
\begin{equation}
\label{eq:screening}
\lambda_i = 
\left\{\begin{array}{cc}
1, &  \rho_i R_i^2 < 3 M \phi_\text{bg}, \\
\approx \dfrac{3 M \phi_\text{bg}}{\rho_i R_i^2}, & \rho_i R_i^2 > 3 M \phi_\text{bg}.
\end{array}\right.
\end{equation}

Consider Fig.~\ref{fig:n1plot} where $n=1$ is fixed and the chameleon force depends on $M$ and $\Lambda$. In region (a), where $M = M_\text{max}$, both the source and test mass are unscreened, i.e. $\lambda_1 = \lambda_2 = 1$. Here the chameleon force depends on $M$ but is independent of $\Lambda$. Assuming the source mass is larger and/or more dense than the test mass it will become screened first, resulting in the transition to region (b). Here $\lambda_1 = 1$ while 
\begin{equation}\label{eq:screeningfactor}
\lambda_2 \approx \frac{3 M \phi_\text{bg}}{\rho_2 R_2^2 }
\end{equation}
where $\rho_{1,2}$ is the density of the objects and
$$
\phi_\text{bg} \equiv \phi_\text{min}(\rho_\text{bg}) = \left( \frac{n M \Lambda^{4+n}}{\rho_\text{bg}} \right)^{1/(n+1)},
$$
where $\rho_\text{bg}$ is the background density. Since $\phi_\text{bg}$ depends on $\Lambda$, the chameleon force is dependent on both $M$ and $\Lambda$ in this region. Further decreasing $M$ or $\Lambda$ will eventually cause the test mass to become screened as well. In region (c), which is defined by $\Lambda = \Lambda_{\alpha}$ where
\begin{equation}\label{eq:Lmin}
\Lambda_\alpha = \left(\frac{F_\text{min}}{4\pi\xi^2 \alpha} \frac{(s+R_1+R_2)^2}{ R_1 R_2}\right)^{3/10} \left(\frac{1}{L}\right)^{2/5} \frac{(\hbar c)^{7/10}}{1.6\times10^{-19}},
\end{equation}
both masses are screened. Here the chameleon force is dependent on $\Lambda$, but not on $M$, as the factor of $M^2$ in Eq. (\ref{eq:force}) is cancelled by the two factors of $M$ from $\lambda_1$ and $\lambda_2$. This phenomena is somewhat special - consider the chameleon force between two objects of fixed size and composition. As the matter-coupling strength of the chameleon is increased, the force will increase until both objects are screened, thereafter, the force is independent of $M$ and remains constant. 
For a finite size vacuum chamber, the background field value $\phi_\text{bg}$ is determined by $n$, $\Lambda$, and $R_\text{vac}$ rather than $\rho_\text{bg}$. However, for sufficiently small $M$ and $\Lambda$, the field will relax back to it's ``natural" minimum, $\phi_\text{bg}$, determined by $\rho_\text{bg}$ which results in the transition to region (d). 

Now consider Fig.~\ref{fig:n>1plot} where $\Lambda = \Lambda_\text{DE}$ is fixed and the chameleon force depends on $M$ and $n$. The qualitative behavior remains the same with $\Lambda$ and $n$ exchanging roles. In this case, region (c), where both masses become screened, is defined by $n = n_\alpha$ where $n_\alpha$ is defined implicitly by
\begin{multline}\label{eq:nmax}
F_\text{min} = 4 \pi \xi^2 \alpha \frac{R_1 R_2}{\left(s+R_1+R_2\right)^2} \left(\frac{1}{\hbar c}\right)^{\dfrac{n_\alpha+6}{n_\alpha+2}} \\
\times \left[n_\alpha(n_\alpha+1)(\Lambda_\text{DE})^{4+n_\alpha}L^2\right]^{\dfrac{2}{n_\alpha+2}}.
\end{multline}

\begin{figure*}
  \includegraphics[]{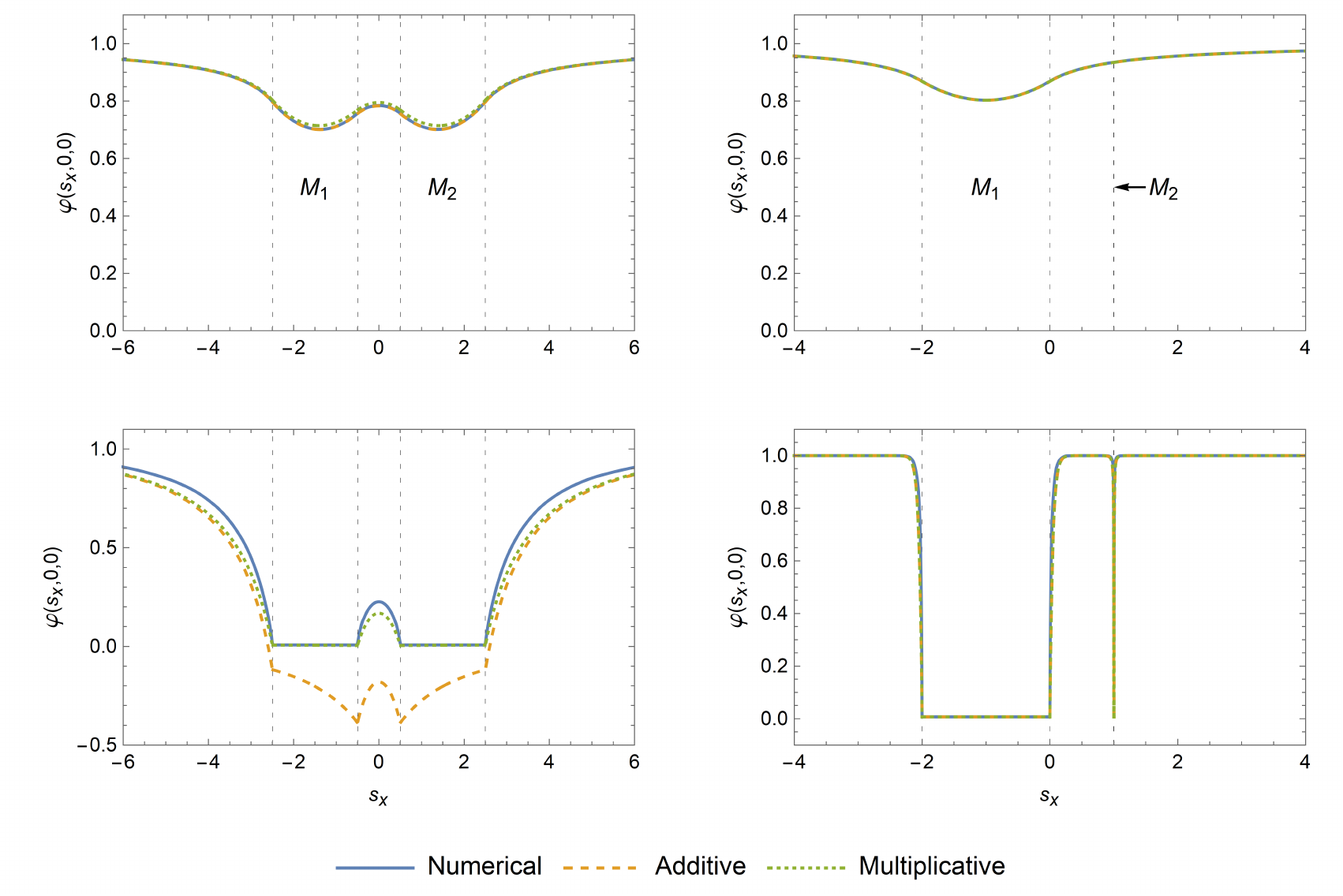}
  \caption{Two-body chameleon field approximations and numerical solution for symmetric spheres in the weakly (top, left) and strongly (bottom, left) perturbing regimes, and asymmetric spheres in the weakly (top, right) and strongly (bottom, right) perturbing regimes. Here $\varphi = \phi/\phi_\text{bg}$ and $s_\text{x} = x/R_1$. Three-dimensional numerical field solutions were found using Mathematica's finite element package. }
  \label{fig:fieldplot}
\end{figure*}

\section{Additive vs. Multiplicative Ansatz}

The chameleon force derived in Refs. \cite{Hui_2009,Burrage_2015} assumes a separation of scales between the source and test masses. This assumption ensures that two requirements are satisfied: first, that second derivatives of the background (source mass) field can be ignored on a scale larger than the test mass, and second, that the field produced by the test mass can be treated as a perturbation to the source mass field. When this mass hierarchy is satisfied, the two-body chameleon field is well approximated by an additive ansatz of the form
$$
\phi(\mathbf{x}) \approx \phi_1(\mathbf{x} - \mathbf{x}_1) + \phi_2(\mathbf{x} - \mathbf{x}_2) - \phi_\text{bg}
$$
where $\phi_{1,2}$ are the one-body chameleon field solutions.

The break down of the additive ansatz depends on a number of variables including sphere size, density, separation distance,
and the chameleon parameters $M$, $\Lambda$ and $n$. When the additive ansatz breaks down, one or both of the requirements for the force derivation in Refs. \cite{Hui_2009,Burrage_2015} are no longer satisfied.

In this regime, we introduce a multiplicative ansatz of the form
$$
\phi(\mathbf{x}) \approx \frac{\phi_1(\mathbf{x} - \mathbf{x}_1) \ \phi_2(\mathbf{x} - \mathbf{x}_2)}{\phi_\text{bg}}.
$$
The utility of this solution rests on the fact that the product will properly approach $\phi_\text{bg}$ far from each sphere, and it will have an appropriate indentation around each sphere without going to unphysical, negative field values.

To investigate and compare the behavior of each ansatz, we solved for the two-body chameleon field using finite element methods. The results are illustrated in Fig. \ref{fig:fieldplot}.

For approximately equal size masses, we find that in the weakly perturbing regime, the additive ansatz provides a better approximation to the exact two-body field solution. Despite this, the force derived using the multiplicative ansatz matches the previous results \cite{Hui_2009,Burrage_2015}. In the strongly perturbing regime, the additive ansatz becomes negative within and between the spheres. Meanwhile, the multiplicative ansatz remains well behaved and provides a good approximation to the numerical solution.

The asymmetric system shown in Fig. \ref{fig:fieldplot} illustrates the two-body field solution for the light blue fill, solid outline microsphere system in Fig. 3 of the main text. For an asymmetric system with a reasonable separation distance, we confirm that the additive ansatz remains valid even in the strongly perturbing regime. Here, the multiplicative ansatz also provides a good approximation to the two-body field solution.

\section{Chameleon Force, Energy Method}

Consider two spheres with masses, $M_{1,2}$, densities, $\rho_{1,2}$, and radii, $R_{1,2}$, centered at $\mathbf{x}_{1,2}$. Our force derivation relies on varying the energy functional of the chameleon field with respect to the position of our spheres similar to the approach in Ref. \cite{Brax_2019}. The chameleon path integral is given by
\begin{equation}\label{eq:path_int}
Z[\rho] = \braket{0 | e^{-i H T} | 0} = \int \mathcal{D}\phi \ e^{i\int d^4x \ \tilde{\mathcal{L}}[\phi,\rho]},
\end{equation}
which, for static sources, yields $Z[\rho]=e^{-i E[\rho] T}$.  Here, the chameleon Lagrangian is given by
$$
\mathcal{L} = -\frac{1}{2}\partial_\mu \phi \partial^\mu \phi - V(\phi) - \frac{\rho}{M}\phi.
$$
Notice that $F=-E[\rho]$ is the field theory analogue of the Helmholtz free energy, $F$, from statistical mechanics. Therefore, the chameleon force on one sphere can be found by varying the energy with respect to the sphere's position,
$$
\mathbf{F}_1 = -\nabla_\text{1}E[\rho].
$$
The energy depends on the position of our spheres through the source,
\begin{multline*}
\rho(\mathbf{x}) = (\rho_1 - \rho_\text{bg}) \theta\left(R_1 - |\mathbf{x} - \mathbf{x}_1|\right) \\
+ (\rho_2 - \rho_\text{bg}) \theta\left(R_2 - |\mathbf{x} - \mathbf{x}_2|\right) + \rho_\text{bg}.
\end{multline*}
Using Eq. (\ref{eq:path_int}),
$$
\nabla_1 E[\rho] = \frac{i}{T} \frac{\int \mathcal{D}\phi \ \left(- i \int d^4x \ \nabla_1 \rho \frac{\phi}{M}\right) e^{i\int d^4x \ \tilde{\mathcal{L}}[\phi,\rho]}}{\int \mathcal{D}\phi \ e^{i\int d^4x \ \tilde{\mathcal{L}}[\phi,\rho]}}
$$
$$
= - \frac{1}{M}\frac{1}{T} \int d^4x \nabla_1 \rho \frac{\int \mathcal{D}\phi \ \phi \ e^{i\int d^4x \ \tilde{\mathcal{L}}[\phi,\rho]}}{\int \mathcal{D}\phi \ e^{i\int d^4x \ \tilde{\mathcal{L}}[\phi,\rho]}}
$$
$$
= \frac{1}{M} \int d^3x \ \nabla_1 \rho \ \phi_\text{cl.}
$$
where $\phi_\text{cl.} = \langle \phi \rangle$ is the classical two-body field configuration. Thus, the chameleon force between the two spheres is given by
$$
\mathbf{F}_1 = - \frac{1}{M} \int d^3x \ \nabla_1 \rho \ \phi_\text{cl.}.
$$
To evaluate the force, we use the multiplicative ansatz as an approximation to the classical field configuration, $\phi_\text{cl.} \approx \phi_1(\mathbf{x} - \mathbf{x}_1) \ \phi_2(\mathbf{x} - \mathbf{x}_2) /\phi_\text{bg}$. Using
$$
\int d^3x \ g(\mathbf{x}) \ \nabla \theta(f(\mathbf{x})) = \int_{f^{-1}(0)} d^2x \  g(\mathbf{x}) \frac{\nabla f(\mathbf{x})}{|\nabla f(\mathbf{x})|},
$$
the force becomes
$$
- \frac{\rho_1 - \rho_\text{bg}}{M \phi_\text{bg}} \int_{|\mathbf{x} - \mathbf{x}_1| = R_1} d^2x \ \left\{ \frac{\mathbf{x} - \mathbf{x}_1}{|\mathbf{x} - \mathbf{x}_1|}  \ \Big[\phi_1(\mathbf{x} - \mathbf{x}_1) \ \phi_2(\mathbf{x} - \mathbf{x}_2)\Big]
\right\}.
$$
The integral can be evaluated using spherical coordinates centered on $\mathbf{x}_1$,
$$
- \frac{\rho_1 - \rho_\text{bg}}{M \phi_\text{bg}}\phi_1(R_1) 2\pi R_1^2 \int_0^\pi \sin{\theta} d\theta \ \left\{\mathbf{\hat{r}}  \ \phi_2(\mathbf{x} - \mathbf{x}_2)
\right\}.
$$
As expected, the $x$ and $y$ components integrate to $0$ and we are left with 
$$
- \frac{\rho_1 - \rho_\text{bg}}{M \phi_\text{bg}}\phi_1(R_1) 2\pi R_1^2 \int_0^\pi \sin{\theta}\cos{\theta} d\theta \ \phi_2(\mathbf{x} - \mathbf{x}_2).
$$
The one-body analytic field solution is given by \cite{Burrage_2015}
$$
\phi_i(\mathbf{x}) = \phi_\text{bg} - \frac{\lambda_i}{4\pi}\frac{M_i}{M} \frac{e^{-m_\text{bg} |\mathbf{x}|}}{|\mathbf{x}|},
$$
where $m_\text{bg}$ is the background chameleon mass. After evaluating the integral, the result is
\begin{multline*}
- \frac{\rho_1 - \rho_\text{bg}}{M \phi_\text{bg}} \phi_1(R_1) \ 2\pi R_1^3 \ \frac{\lambda_2}{4\pi}\frac{M_2}{M} \ \times \\
\frac{2}{(m_\text{bg}R_1)^3} \frac{1 + m_\text{bg}d}{d^2} \left[m_\text{bg}R_1 \cosh{m_\text{bg}R_1} - \sinh{m_\text{bg}R_1} \right] e^{-m_\text{bg}d}
\end{multline*}
where $d$ is the distance between the spheres. For the cases of interest, $\rho_\text{bg} \ll \rho_1$, and the force becomes
\begin{multline*}
F_1(d) = - \frac{\phi_1(R_1)}{\phi_\text{bg}} \frac{\lambda_2}{4\pi} \ \frac{M_1 M_2}{M^2} \ \times \\
\frac{3}{(m_\text{bg}R_1)^3} \frac{1 + m_\text{bg}d}{d^2} \left[m_\text{bg}R_1 \cosh{m_\text{bg}R_1} - \sinh{m_\text{bg}R_1} \right] e^{-m_\text{bg}d}.
\end{multline*}
We find a chameleon force that has $\sinh$ and $\cosh$ factors similar to the result found in Ref. \cite{Qvarfort_2021}. In the limit $m_\text{bg}R_1 \ll 1$, which is satisfied for the cases of interest, 
$$
F_1(d) \approx - \frac{\phi_1(R_1)}{\phi_\text{bg}} \frac{\lambda_2}{4\pi} \ \frac{M_1 M_2}{M^2} \
\frac{1 + m_\text{bg}d}{d^2} e^{-m_\text{bg}d}.
$$
If we further take $m_\text{bg}d \ll 1$, we find that
\begin{equation}
F_1(d) \approx - \frac{\phi_1(R_1)}{\phi_\text{bg}} \frac{\lambda_2}{4\pi} \ \frac{M_1 M_2}{M^2} \
\frac{1}{d^2}.
\end{equation}
In order to evaluate $\phi_i(R_i)$, recall that the screening factor $\lambda_i$ is given by \cite{Burrage_2015}
$$
\lambda_i \approx \left\{
\begin{array}{cc}
1 & \dfrac{2}{3}\dfrac{M \phi_\text{bg}}{\frac{1}{4\pi}\frac{M_i}{R_i}} \geq 1\\
1 - \left(\dfrac{S_i}{R_i}\right)^3 & \dfrac{2}{3}\dfrac{M \phi_\text{bg}}{\frac{1}{4\pi}\frac{M_i}{R_i}} \leq 1
\end{array}\right.
$$
where
$$
S_i \approx \sqrt{1 - \frac{2}{3}\frac{M \phi_\text{bg}}{\frac{1}{4\pi}\frac{M_i}{R_i}}}
$$
in the limit $m_\text{bg}R_i \ll 1$. In the weakly perturbing regime, 
$$
\frac{\phi_1(R_1)}{\phi_\text{bg}} \approx 1.
$$
In the strongly perturbing regime, $\phi_1(R_1)/\phi_\text{bg} \approx 0$ to first order in $\lambda_1$. Expanding $\lambda_1$ to second order we find
$$
\frac{\phi_1(R_1)}{\phi_\text{bg}} \approx \frac{\lambda_1}{6}.
$$
where $\lambda_1$ is given by Eq. (\ref{eq:screeningfactor}). When the test mass (sphere 1) is weakly perturbing, our force expression agrees with Eq. (\ref{eq:force}) with $\alpha = 1$ \cite{Burrage_2015,Hui_2009}. However, when the test mass is strongly perturbing our force differs by a factor of $\alpha = 1/6$.

\section{``Safe" Chameleon Parameter Space}

At early cosmological times prior to Big Bang nucleosynthesis (BBN), the chameleon remains light and gets stuck along its potential due to Hubble friction \cite{Brax_2004}. However, BBN requires that the chameleon reach its minimum prior to the onset of nucleosynthesis. It was shown in Ref. \cite{Brax_2004} that as particle species become nonrelativistic, the chameleon experiences several kicks, pushing it toward the minimum of its potential. Fortunately, the kicks allow the chameleon to reach the minimum of its potential prior to BBN for a wide range of initial conditions.

However, in Refs. \cite{Erickcek_2013,Erickcek_2014} it was shown that for strongly coupled chameleons, these kicks cause the chameleon to follow the so-called ``surfing" solution characterized by a constant Jordan-frame temperature. Surfing chameleons have sufficient kinetic energy when they reach the potential minimum that they are able to climb the steep side of the potential. This leads to rapid changes in the chameleon mass which results in extremely high energy particle production. These high energy modes undermine the treatment of the chameleon as an effective field theory and pose a danger to BBN. In Ref. \cite{Erickcek_2014} it was shown that chameleons with $M/M_\text{p} \lesssim 1/1.82$ are able to surf, defining a safe region of parameter space for greater $M$. However, targeting smaller $M$ can still place constraints on modified chameleon theories such as the Dirac-Born-Infeld (DBI) chameleon \cite{Padilla_2016}.

On the other hand, there is also a limit on how weakly coupled the chameleon can be if we wish to use it as a quintessence field. The existence of an attractor solution for the chameleon field was shown in Ref. \cite{Brax_2004}. Along the attractor, the chameleon will follow the minimum of its potential provided $m^2/H^2 > 1$ where 
\begin{equation}\label{eq:chameleonmass}
m^2 = m^2(\rho) = \frac{n(n+1)\Lambda^{4+n}}{\phi_\text{min}^{n+2}(\rho)}
\end{equation}
is the chameleon mass and $H$ is the Hubble parameter. This inequality can be used to place an upper bound on $M$. In the limit $\phi \lesssim \Lambda$, \cite{Brax_2004}
$$
\frac{m^2}{H^2} > 3 (1+2n) \frac{M_\text{p}^2}{M \Lambda} \Omega_\text{m},
$$
where $\Omega_\text{m}$ is the matter density parameter. Requiring that this be greater than $1$ yields
$$
\frac{M}{M_\text{p}} < 3 (1+2n) \frac{M_\text{p}}{\Lambda} \Omega_\text{m}.
$$
Since $\Omega_\text{m}$ is monotonically increasing from $\Omega_\text{m} \sim 10^{-28}$ \cite{Brax_2004}, this provides a maximum bound for $M$. In the limit $\phi \gg \Lambda$ this yields \cite{Brax_2004}
$$
\frac{m^2}{H^2} \approx 3 \frac{M_\text{p}^2}{M \Lambda} (n+1)\frac{\Lambda}{\phi}\Omega_\text{m}.
$$
Using
$$
\frac{\Lambda}{\phi} \sim \left(\frac{1}{n}\frac{\Lambda}{M}\Omega_\text{m}\frac{3 H^2 M_\text{p}^2}{\Lambda^4}\right)^{1/(n+1)},
$$
the inequality bounding $M$ becomes
\begin{multline*}
M \leq \left[3(n+1)M_\text{p}^{(n+2)/(n+1)}\left(\frac{M_\text{p}}{\Lambda}\right)^{n/(n+1)} \right. \\ \left. \Omega_\text{m}^{(n+2)/(n+1)}\left(\frac{1}{n}\right)^{1/(n+1)}\left(\frac{3 H^2 M_\text{p}^2}{\Lambda^4}\right)^{1/(n+1)}\right]^{(n+1)/(n+2)}.
\end{multline*}
Since $H(z) \geq H_0$ and $\Omega_\text{m} \gg 10^{-6}$ for $\phi \gg \Lambda$ \cite{Brax_2004}, the bound on $M$ follows from setting $H \to H_0$ and $\Omega_\text{m} \to 10^{-6}$. This provides an upper bound on $M$ for the yellow region in Fig. 4 of the main text.

Additionally, in Ref.~\cite{Upadhye_2012} a bound on the maximum chameleon mass was found by requiring that quantum corrections to the potential energy remain small. This conditions provides another minimum bound on $M$, but from a quantum field theory perspective as opposed to the cosmological implications explored in Ref. \cite{Erickcek_2014}. From Ref. \cite{Upadhye_2012} the inequality constraining the chameleon mass is
$$
m(\rho) \leq \left(\frac{48\pi^2\rho^2}{M^2}\right)^{1/6}
$$
where the chameleon mass is given by \ref{eq:chameleonmass}. This yields an implicit bound on $M$ as a function of $\Lambda$ and $n$ illustrated by the blue region in Fig. 4 of the main text.


\bibliographystyle{apsrev4-1}
\bibliography{supplemental}